\documentclass[conference]{IEEEtran}
\IEEEoverridecommandlockouts

\usepackage{cite}
\usepackage{amsmath,amssymb,amsfonts}
\usepackage{algorithmic}
\usepackage{graphicx}
\usepackage{textcomp}
\usepackage{xcolor}
\usepackage{tabularx} 

\usepackage{mdwlist}
\usepackage[ruled,vlined,shortend, linesnumbered]{algorithm2e} 
\usepackage[]{algorithmic} 
\usepackage{booktabs}
\usepackage{color}

\usepackage{amsthm,amssymb}
\usepackage{amsmath}
\usepackage[font=small]{caption}
\usepackage[skip=3pt]{subcaption}
\usepackage{multirow}
\usepackage{mathrsfs}
\usepackage{amsbsy}
\usepackage[mathscr]{eucal} 
\usepackage[flushleft]{threeparttable}
\usepackage[hidelinks]{hyperref}
\usepackage{comment}
\usepackage{paralist}
\usepackage{pifont}

\usepackage{bbding} 

\usepackage{svg}

\usepackage{dsfont}

\def\BibTeX{{\rm B\kern-.05em{\sc i\kern-.025em b}\kern-.08em
    T\kern-.1667em\lower.7ex\hbox{E}\kern-.125emX}}

\newcommand{\hide}[1]{}

\newcommand{\bit}{\begin{compactitem}}
\newcommand{\eit}{\end{compactitem}}
\newcommand{\ben}{\begin{compactenum}}
\newcommand{\een}{\end{compactenum}}

\begin{document}

\newcommand{\cindy}[1]{\textcolor{blue}{#1}}

\newcommand{\tina}[1]{\textcolor{green}{#1}}

\title{Sparsity-exploiting Gaussian Process for Robust Transient Learning of Power System Dynamics 
}

\author{\IEEEauthorblockN{Tina Gao}
\IEEEauthorblockA{Electrical and Computer Engineering \\
Carnegie Mellon University}
\and
\IEEEauthorblockN{Shimiao Li}
\IEEEauthorblockA{Electrical Engineering \\
University at Buffalo}
\and
\IEEEauthorblockN{Lawrence Pileggi}
\IEEEauthorblockA{Electrical and Computer Engineering \\
Carnegie Mellon University}

\thanks{This manuscript has been submitted to PESGM2026.
\copyright 2021 IEEE. Personal use of this material is permitted. Permission from IEEE must be obtained for all other uses, in any current or future media, including reprinting/republishing this material for advertising or promotional purposes, creating new collective works, for resale or redistribution to servers or lists, or reuse of any copyrighted component of this work in other works.}
}

\maketitle

\begin{abstract}
Advances in leveraging Gaussian processes (GP) have enabled learning and inferring dynamic grid behavior from scarce PMU measurements. However, real measurements can be corrupted by various random and targeted threats, leading to inaccurate and meaningless results. This paper develops robust transient learning to overcome this challenge by exploiting the sparse corruption patterns in the data flow. Specifically, we integrate sparse optimization with method of moments (MoM) to make learning robust to a sparse distribution of data corruptions; then, we optimize sparse weights to identify corrupted meter locations. To improve inference speed on large-scale systems, we further adopt K-medoid clustering of locations to develop dimension reduction (DR) and aggregate representation (AR) heuristics. Experimental results demonstrate robustness against random large errors, targeted false data injections, and local PMU clock drifts. On a 1354-bus system, inference turns out to be 18x faster using DR and 400x faster when further combined with AR heuristics.

\end{abstract}


\section{Introduction}
\label{sec:Introduction}

Safeguarding the electrical power grid against transient instability risks is of vital importance and can be enhanced by closely monitoring the phase angles, frequencies, and rates of change of frequency (ROCOF) from phasor measurement units (PMUs) \cite{Standard-PMU-IEEE}. However, due to the sparsity of PMU installations \cite{pmu_coverage2022} in practice, full observability of system dynamics is not available, which motivates various transient learning techniques to infer dynamic behavior from scarce measurement readings.
Many existing methods such as matrix completion, principal component analysis, and tensor decomposition techniques are model-free, aiming to reconstruct missing PMU data but are limited to filling the missing signals at locations where PMUs are already installed. In contrast, recent advances leverage Gaussian processes (GP) \cite{GP-PMU-Vassilis} that incorporate physical models of swing equations. When extended to a multi-input multi-output (MIMO) setting, GP can infer dynamics at non-measured non-PMU locations, allowing transient insights more widely across the network \cite{GP-PMU-Vassilis}.

\begin{figure}[ht]
     \centering
     \begin{subfigure}[h]{\linewidth}
         \centering         
         \includegraphics[height=0.7in]{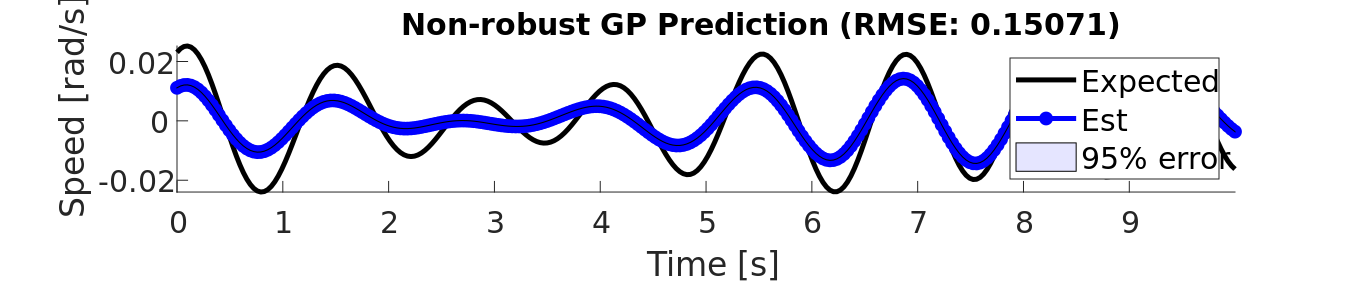}
     \end{subfigure}
     \begin{subfigure}[h]{\linewidth}
         \centering         
         \includegraphics[height=0.7in]{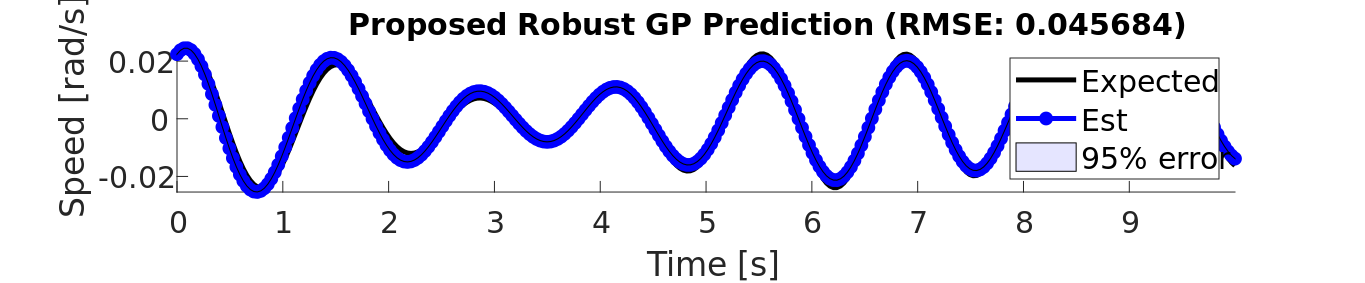}
     \end{subfigure}
      \caption{{300-bus case with false data injections}: Dynamic behavior is predicted for non-metered location in the presence of false data injections. Traditional GP is inaccurate (top), whereas result from our robust method remains accurate (bottom).}
      \label{fdiaEst}
\end{figure}

Although synchrophasor data and transient learning techniques have provided unprecedented opportunities to identify transient instability risks, the evolving cyberthreat landscape presents growing challenges in accurately learning the dynamics of the power system from real-world data. In practice, PMU data are susceptible to random errors from various sources, including communication noise, local clock drift, synchronization delays, etc. Adding to these random disturbances is a new adversary in the form of cyber-intrusions, capable of causing interactive bad data and even well-designed manipulations. As studied in \cite{gridsecurity-Vyas}\cite{PMU_vulnerable_GPSjamming}\cite{PMU_vulnerable_GPSspoofing}\cite{PMU_FDIA}, malicious agents can exploit multiple vulnerabilities specific to PMU to significantly compromise data integrity, creating opportunities for jamming\cite{PMU_vulnerable_GPSjamming} (attackers cause a PMU to lose track of the GPS synchronization signal), spoofing\cite{PMU_vulnerable_GPSspoofing} (attackers inject fake GPS signals), and false data injections\cite{PMU_FDIA} (attackers modify PMU data). 
As an illustrative example, the top plot in Figure \ref{fdiaEst} visually indicates that the state-of-the-art GP method is not robust to these risks of data corruption, resulting in bad (and almost meaningless) predictions for non-metered generators. 

\begin{figure}[ht]
  \subcaptionbox*{}[.5\linewidth]{%
    \includegraphics[width=\linewidth]{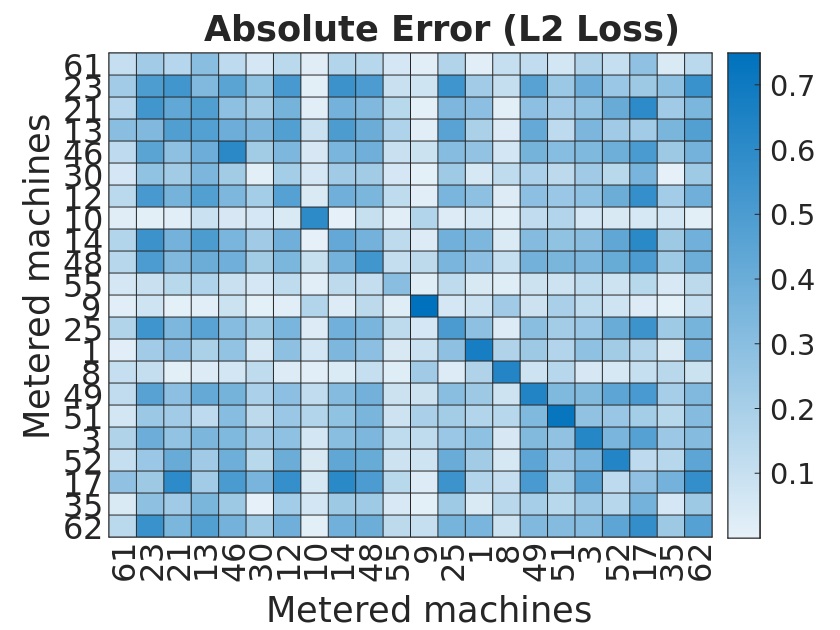}%
  }%
  \hfill
  \subcaptionbox*{}[.5\linewidth]{%
    \includegraphics[width=\linewidth]{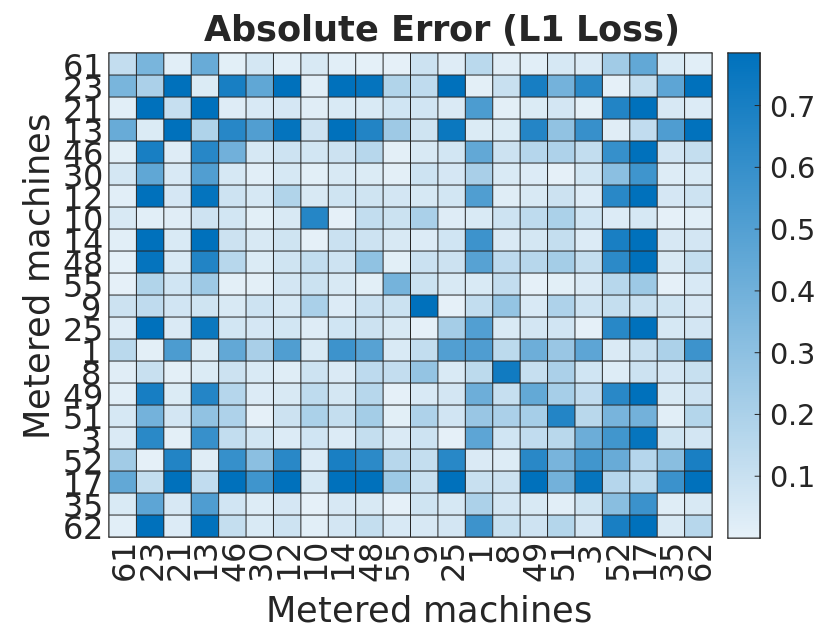}%
  }
  \caption{Exploiting sparse data-corruption patterns: The difference between learned and sample correlation matrices is plotted when 5 PMUs are corrupted among a total of 22 meters. Our proposed method (right) \lq{pushes}\rq\, large deviations to appear at sparse rows and columns corresponding to bad locations, accurately learning the true covariance in which bad locations are \lq{\lq{correctified}\rq}\rq. The classic GP learning (left) cannot capture this structure, and the learned correlation is corrupted almost everywhere by the falsified data. }
  \label{sparseCovPattern}
\end{figure}

To address this concern, this paper proposes robust transient learning to accurately capture the damping and oscillatory behavior when inaccurate measurements that distort observations of system dynamics are encountered. We exploit the sparse PMU data corruption patterns to this end. This is motivated by state-of-the-art robust state/parameter estimation approaches\cite{sugar-robustSE-Li}\cite{sugar-robustGSE-Li} that identify and reject sparse bad data from a state-state snapshot-based set of measurements; and this paper extends robust data analytics to the time-series transient horizon, based on a similar assumption that the distribution of corrupted data is sparse across locations and/or times. Specifically in this work, when the $k$-th PMU is bad among a total of $m$ installed PMUs, the sample data covariance matrix ($m\times m$ matrix calculated from time-series measurements collected over $T$ time ticks) is observed to deviate largely from the true covariance at the $k$-th row and column, while the remaining entries remain aligned. Leveraging this sparse structure, we develop a two-step process. First, we make learning immune to the sparse distribution of corrupted meters using sparse optimization that minimizes a least absolute value (L1-norm) based objective in the Method of Moments (MoM) based learning process. As a result, our method captures sparse deviations on the covariance (see Figure \ref{sparseCovPattern} right), whereas the non-robust GP counterpart fails to leverage this sparse structure (see Figure \ref{sparseCovPattern} left).
In the second step, we further develop a sparse auto-masking method that extends the robust learning with a sparse weight vector to explicitly indicate the corrupted meter locations and discard data from corrupted meters when making predictions. 
For example in Figure \ref{fdiaEst}, the bottom plot illustrates the efficacy of our two-step methodology for accurately predicting dynamic behaviors at non-metered locations (identified corrupted data are discarded).

Finally, this paper extends the use of the method to large systems. We present a correlation-based clustering method to sparsify a covariance matrix for faster computation. Although this clustering method is analogous to the workflow for coherency identification \cite{Cluster-Coherency-Zoupash}, its application here is uniquely efficient because our parameterized covariance framework allows us to interrogate necessary correlation entries without simulating the entire system.

\textbf{Reproducibility:} our code and data are publicly available at: {\color{blue} https://github.com/tinagaostrawberry/Sparse-Gaussian-Process-Power-System-Dynamics}.

\section{Background}
\label{sec:Background}

\subsection{Gaussian Processes for Linear Time Invariant Systems}

A Gaussian process (GP)\cite{GP-ML-Rasmussen} models a set of random variables whose finite subsets follow a joint Gaussian distribution. When modeling a time-varying function space ${x(t)}$, the joint distribution of two signals ${\mathbf{x_1} = x(\mathcal{T}_1)}$ and ${\mathbf{x_2} = x(\mathcal{T}_2)}$, where ${\mathcal{T}_1}$, ${\mathcal{T}_2}$ are two sets of time ticks, is:
\begin{equation}\label{eqnGpJoing}
\begin{bmatrix}
\mathbf{x_1} \\
\mathbf{x_2} \\
\end{bmatrix}
\sim
\mathcal{N}(
\begin{bmatrix}
\mathbf{\mu_1} \\
\mathbf{\mu_2} \\
\end{bmatrix},
\begin{bmatrix}
\mathbf{\Sigma_{11} \Sigma_{21}}^\top \\
\mathbf{\Sigma_{21} \Sigma_{22}}^\top
\end{bmatrix}
)
\end{equation}
In this framework, given an observed ${\mathbf{x_1}}$, inferring the unobserved ${\mathbf{x_2}}$ can be obtained from the conditional probability function of ${\mathbf{x_2}}$ given ${x_1}$:
\begin{subequations}\label{eqnGpMmse}
\begin{align}
\mathrm{E}[\mathbf{x_2}|\mathbf{x_1}] = 
\mathbf{\mathrm{\mu}_2} + 
\mathbf{\mathrm{\Sigma}_{21}\mathrm{\Sigma}_{11}}^{-1}
(\mathbf{x_1}-\mathbf{\mathrm{\mu}_1)}
\label{eqnGpMmse:a}
\\
\mathrm{Cov}[\mathbf{x_2}|\mathbf{x_1}] =
\mathbf{\mathrm{\Sigma}_{22}} - 
\mathbf{\mathrm{\Sigma}_{21}
\mathrm{\Sigma}_{11}}^{-1}
\mathbf{\mathrm{\Sigma}_{21}}^\top
\label{eqnGpMmse:b}
\end{align}
\end{subequations}

To acquire the covariance matrices, covariances can be parameterized by a kernel function that is typically learned through maximum likelihood estimation.

\subsection{GP Application to MIMO Power System Dynamics}


Power grid dynamic behavior can be modeled with simplicity using linearized swing equations at a network level:
\begin{equation}\label{eqnSysSwing}
\mathbf{M\dot{\omega}(t)} + \mathbf{D\omega(t)} + \mathbf{L\theta(t)} = \mathbf{p(t)}
\end{equation}
where ${\mathbf{\omega(t)}}$ and ${\mathbf{\theta(t)}}$ are column vectors of speed and rotor angles from all system generators comprising set ${\mathcal{N}}$. ${\mathbf{p(t)}}$ is the vector of electric-mechanical power mismatch, ${\mathbf{M}}$ and ${\mathbf{D}}$ are diagonal matrices of generator inertia and damping coefficients, and ${\mathbf{L}}$ is the negative Jacobian matrix of power flow equations after Kron reduction.

  Recent advances of GP \cite{GP-PMU-Vassilis} assume uniform damping and exploit the transformation between the above multi-input multi-output (MIMO) system and a set of single-input-single-output (SISO) eigensystems, via eigen decomposition. Each SISO eigensystem is described as $\ddot{y}(t) + \gamma \dot{y}(t) + \Lambda y(t) = x(t)$
where ${\mathbf{x(t)} := \mathbf{V}^\top \mathbf{M}^{-1/2} \mathbf{p(t)}}$ is a vector of eigeninputs, ${\mathbf{y(t)} := \mathbf{V}^\top \mathbf{M}^{1/2} \mathbf{\theta (t)}}$ is a vector of eigenstates, and ${\mathbf{V}}$ and ${\mathbf{M}}$ are eigenvector and eigenvalue matrices of ${\mathbf{M}^{-1/2}\mathbf{LM}^{-1/2}}$.

Now, let us define the problem as inferring grid dynamics using GP, e.g. inferring generator speeds ${\mathbf{w(t)}}$ at non-metered buses, using data collected from a set of PMU-installed generators  ${\mathcal{M}}$. This requires learning the covariance matrix $\mathrm{E}[\mathbf{\omega(t+\tau) \omega(t)}^\top]$, which, according to the above MIMO-SISO transformation, can be further reduced to learning the covariance ${\mathrm{E}[\dot{y}(t+\tau) \dot{y}(t)^\top] }$. Let ${\mathrm{E}[\dot{y}(t+\tau) \dot{y}(t)^\top] }$ be parameterized by $\mathbf{A}$, then the task is to learn $\mathbf{A}$ and calculate by
\begin{equation}\label{eqnCovSpeed}
\mathrm{E}[\mathbf{\omega(t+\tau) \omega(t)}^\top] =
\mathbf{M}^{-1/2} \mathbf{V}
(\mathbf{A} \odot \mathbf{K_{\tau}})
\mathbf{V}^\top \mathbf{M}^{-1/2}
\end{equation}
\noindent where ${\odot}$ denotes element-wise multiplication and ${\mathbf{K_{\tau}}}$ is a custom spatio-temporal kernel facilitating the parameterization of ${\mathrm{E}[\dot{y}(t+\tau) \dot{y}(t)^\top] }$ by ${\mathbf{A}}$. The full definitions and derivation, as well as an optional model reduction step that reduces the size of ${\mathbf{A}}$ through bandpass filtering, is detailed in \cite{GP-PMU-Vassilis}.

To obtain the optimal $\mathbf{A}$, standard GP leverages method-of-moments \cite{MOM-Kay} which is a least-squares method non-robust to corrupted data:
\begin{equation}\label{eqnLearn}
\min_{\mathbf{A}}
\quad ||vectorize(\mathbf{\Sigma}_{\tau}(\mathbf{A}) - \mathbf{C}_{\tau})||_2^2
\end{equation}
where ${\mathbf{C_{\tau}} = \frac{1}{T} \sum_{t} \mathbf{z(t+\tau) z(t)}^{\top}}$ is the sample covariance matrix and ${\mathbf{\Sigma}_{\tau}(\mathbf{A}) =
\mathbf{S_{\mathcal{M}} M}^{-1/2} \mathbf{V}
(\mathbf{A} \odot \mathbf{K_{\tau}})
\mathbf{V}^{\top} \mathbf{M}^{-1/2} \mathbf{S_{\mathcal{M}}}^{\top}}$ is the parameterized covariance matrix to be learned. ${T}$ is the number of time ticks, ${\mathbf{S_{\mathcal{M}}}}$ is a matrix that selects ${\mathbf{\omega(t)}}$ in ${\mathcal{M}}$, and ${\mathbf{z(t)}=\mathbf{S_{\mathcal{M}} \omega(t)}}$ is the measurements from metered buses. Subsequently, the estimated ${\mathbf{A}}$ is used to predict non-metered speeds using \eqref{eqnCovSpeed} and \eqref{eqnGpMmse:a}.

\section{Method}
\label{sec:Method}
Let us suppose a total of $n$ locations of interest on a power grid, among which $m$ (generator) locations are installed with PMUs, with $k$ PMUs corrupted. The PMUs collect observations over $T$ time ticks, giving multivariate time-series data ($T\times m$). Our goal is to learn and infer power system dynamic behaviors from these data, accurately and robustly. A two-step learning (training) process is developed to this end:

\begin{enumerate}
    \item \textbf{Robust covariance learning:} Model and learn the covariance matrix with robustness against a sparse distribution of corruptions over time. 
    \item \textbf{Corrupted Meter Identification:} Explicitly identify the locations of corrupted PMUs so that we can discard bad data from corrupt meters when making inferences using GP and address bad meters in reality. 
\end{enumerate}
After learning, we perform inference using GP in \eqref{eqnGpMmse:a}. 

Furthermore, the bottleneck in the above equation is evidently the inversion of ${\Sigma_{11}}$, and, despite the applicability of matrix inversion lemmas \cite{GP-PMU-Vassilis}, it remains a computational burden. Section \ref{sec: clustering for large sysetm} introduces clustering-based heuristics, inspired by a coherency identification method \cite{Cluster-Coherency-Zoupash}, to group generators that exhibit similar dynamics and reduce computation complexity.

\subsection{Robust Learning of Covariance}\label{sec: robust cov learning}
Let $\mathbf{C_{\tau}}$ denote sample covariance calculated from measured data, and $\Sigma_{\tau}$ denote the covariance matrix to be learned and used to make inference by GP. $\Sigma_{\tau}$ can be modeled as a function of learnable parameter $A$, as illustrated in (\ref{eqnLearn}). Using Method of Moments (MoM), the classic GP learns model parameters $A$ by minimizing least square errors between ${C_{\tau}}$ and $\mathbf{\Sigma_{\tau}(A)}$, i.e., minimizing L2 norm $\min_\mathbf{A} ||vectorize(\mathbf{\Sigma_{\tau}(A)}-\mathbf{C_{\tau}})||_2$. The use of least squares or L2 norm as the objective function makes the learning non-robust to data corruptions that incur large deviations from the true values. 

In this Section, we propose to make the MoM learning robust to sparse corruptions, allowing for sparse large errors between least absolute value (L1 norm) based objective, modifying \eqref{eqnLearn} to allow for sparse differences between entries of the learned and sample covariances:
\begin{equation}\label{eqnLoss}
\min_\mathbf{A}
||vectorize(\mathbf{\Sigma_{\tau}(A)}-\mathbf{C_{\tau}})||_1
\end{equation}



During learning, we use correlation in lieu of covariance. The inherent sampling noise is proportional to the magnitude of the data stream, rendering it difficult to distinguish bad data from the sampling noise of larger-magnitude data. Dividing the covariance by its standard deviations normalizes the data and its sampling noise, mitigating the tendency for larger data to dominate the sparse error characterization. However, for simplicity, we slightly abuse the notation in our equations and refer to all correlation matrices as covariances, since the un-normalized correlation matrix is equivalent to the covariance matrix, and the correlation matrix is converted to the covariance matrix in all GPs calculations anyway.

\subsection{Identification and Mitigation of Corrupt Meters}\label{sec: sparse weighting}

Let  $\mathbf{w}=[w_1, w_2,..., w_m]^\top$ be a vector of binary weights to $m$ PMU locations to denote whether or not each PMU is a bad or not, with zero weight $w_i=0, \forall i$ denotes a good meter, and else denotes corrupted. This sparse weights can then be mapped to a weight matrix $\mathbf{W}$ pointing out the covariance entries affected by corrupted data:
\begin{equation}
    \mathbf{W} = \mathbf{w}\mathbf{w}^\top + \mathbf{w}\mathbf{(1-w)}^\top + \mathbf{(1-w)}\mathbf{w}^\top
\end{equation}

Taking a toy example of 5 PMUs in total with 3rd PMU being bad, we ideally return $\mathbf{w}=[0,0,1,0,0]^\top$, and weight matrix  
\begin{equation}
    \mathbf{W} = \begin{bmatrix}
    0       & 0 & 1 & 0 & 0 \\
    0       & 0 & 1 & 0 & 0 \\
    1       & 1 & 1 & 1 & 1\\
    0       & 0 & 1 & 0 & 0\\
    0       & 0 & 1 & 0 & 0
\end{bmatrix}
\end{equation}

With the weights defined above, we can extend the sparse optimization in \eqref{eqnLoss} to further identify bad PMUs.  Also leveraging the sparse distribution of bad meters, the weight vector is enforced to be sparse. We solve an optimization as below:
\begin{subequations}\label{eqnMask}
\begin{align}
\min_\mathbf{w}
\quad & ||vectorize((\mathbf{\Sigma_{\tau}(A)}-\mathbf{C_{\tau}}) \odot \mathbf{M})||_1
+ \beta||\mathbf{w}||_1 
\label{eqnMask:a}
\\
\textrm{s.t. }
\quad & \mathbf{M} = \mathds{1}_{m \times m} - 
(\mathbf{w}\mathbf{w}^\top + \mathbf{w}\mathbf{(1-w)}^\top+\mathbf{(1-w)}\mathbf{w}^\top)
\label{eqnMask:b}
\\
\quad & 0 \leq w_i \leq 1, \forall i
\label{eqnMask:c}
\end{align}
\end{subequations}
where $\mathbf{M}=1-\mathbf{W}$ is a mask matrix that overlays the correlation matrix errors, ideally as a binary mask to selectively exclude sparse sets of rows and columns corresponding to corrupt meters. The L1 regularization term ${||\mathbf{w}||_1}$ encourages the number of corrupt meters to be sparse.

After we learn $\mathbf{w}$, each $w_i$ is binarized with a threshold of 0.5 and is used to indicate the set of corrupted meters. Consequently, during the inference of future or non-metered generators, predictions are only conditioned on valid data (whose $w_i=0$).

Identifying corrupt meters can be difficult when the corruption error is small, either due to the inherently small nature of the noise or the existence of noise only at a few time samples. When it is necessary to increase the sensitivity to corrupted data, we can learn covariances with respect to a larger set of lags ${\tau}$, increasing the opportunities to raise the prominence of noise at corrupt meters. 
This results in the following modifications to \eqref{eqnLoss} and \eqref{eqnMask:a}, respectively.

\begin{equation}\label{eqnLossLag}
\min_{\mathbf{A}} 
\sum_{\tau \in T}
||vectorize(\mathbf{\Sigma_{\tau}(A)} - \mathbf{C_{\tau}})||_1
\end{equation}
\begin{equation}\label{eqnMaskLag}
\min_{\mathbf{w}}
\sum_{\tau \in T}
||vectorize((\mathbf{\Sigma_{\tau}(A)} - \mathbf{C_{\tau}}) \odot \mathbf{M})||_1
+ \beta||\mathbf{w}||_1 
\end{equation}

\subsection{Inference Speed-up for Large Systems}\label{sec: clustering for large sysetm}

The learned covariance model from data provides a measure of correlation between generator locations, without having to simulate the entire system. We leverage this correlation to perform a k-medoids clustering, grouping generators with similar/correlated behaviors together. 

The clustering then enables a two-fold speedup: 1) {\bf dimension reduction}: the inference of each non-metered generator is only conditioned on metered machines within the same clustering group, reducing dimension of covariance of interest by discarding rows and columns of metered machines with weak correlations, circumventing the need to invert the full matrix; 2) {\bf aggregate representation}: machines in the same cluster can be jointly modeled as a signal equivalent generator (approximately), reducing the dimension of the model. 

\section{Experiments}
\label{sec:Experiments}

In the following three subsections, we compare with non-robust GP baseline to demonstrate the efficacy of our method on different sized systems (30-bus, 300-bus, and 1354-bus systems). We evaluated 3 types of corruption: targeted false data injection, large random measurement errors, and local clock drift. 
For false data injection attack and local clock drift, we apply a bandpass filter for model reduction, but for large random errors, we forgo the model-reduction filter, which would substantially curb errors. Experiment settings for each IEEE standard transmission grid test cases are illustrated below:
\begin{itemize}
\item Case30 (6 generators): To illustrate unfiltered oscillations, we opt for a higher reporting rate of 240 samples persecond and a total timeframe of 10 seconds. We set ${\beta=0.8}$.
\item IEEE 300-bus case (69 generators): We
apply a model reduction filter with a [0.5,0.8] Hz bandpass \cite{GP-PMU-Vassilis}. Data  reporting rate is 30 samples per second in a timeframe of 375 seconds. We set ${\beta=6}$.
\item 1354-bus pegase case (260 generators): We apply a model-reducing filter with a [1,1.75] Hz bandpass, use reporting rate of 50 samples per second, set the timeframe to 5.5 seconds, let ${\beta=50}$, and generate data at a timestep of 0.1 ms. Among the 99 metered generators randomly selected, we inject false data onto 10 generators using the same process applied to the IEEE 300-bus case.
\end{itemize}
All tests use synthetic data, and the sets of metered and non-metered machine(s) are randomly selected. The reporting rate and timeframe is selected in each case to collect a sufficiently large number of sample times to ensure the sample covariance closely converges to the true covariance, which is especially salient in identifying corruptive noise. 


The normalized root mean square error (RMSE) is utilized as an accuracy metric:
\begin{equation}\label{nomralizedRmse}
RMSE_{norm} = \frac
{\sqrt{|| \sum_t w_{est}(t)-w_{actual}(t)||_2^2
/T}}{w_{actual,MAX} – w_{actual,MIN}}
\end{equation}

\subsection{False Data Injection}

We randomly select a subset of PMUs to insert false data generated from a perturbed power input ${p(t)}$ and load values. 
As shown in Figure \ref{sparseCovPattern}, our robust learning captures the 5 corrupted meters from the sparse error structure.  Figure \ref{fdiaEst} demonstrated the improved robustness when using our method to predict trasient behaviors at non-metered locations.

\subsection{Large Random Error}
\begin{figure}[h]
     \centering
     \begin{subfigure}[h]{\linewidth}
         \centering
         \includegraphics[width=1\linewidth]{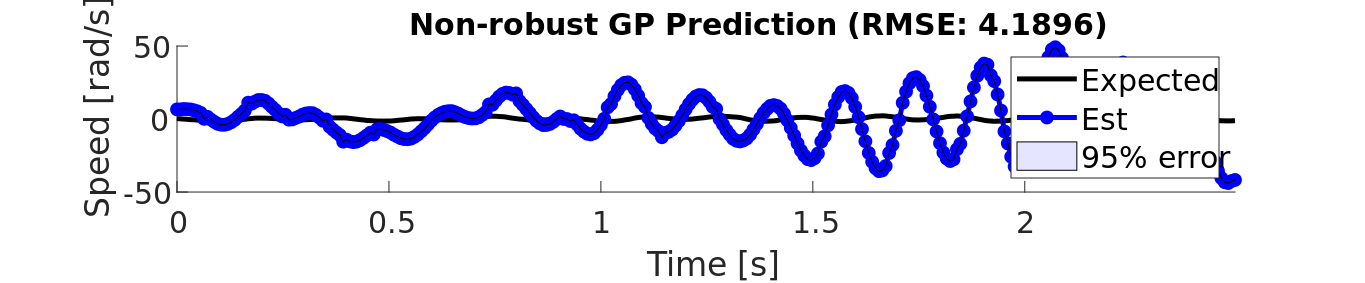}
     \end{subfigure}
     \begin{subfigure}[h]{\linewidth}
         \centering
         \includegraphics[width=1\linewidth]{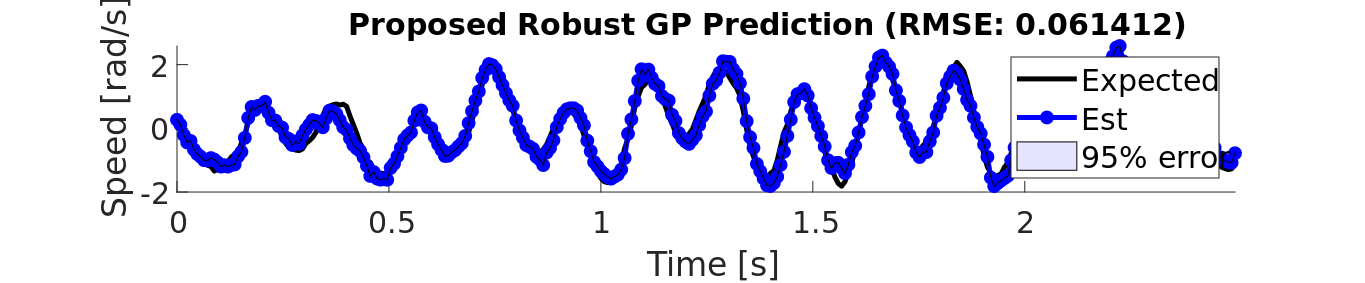}
     \end{subfigure}
      \caption{{30-bus case with random large errors}: dynamic behavior is predicted at non-metered location. Non-robust GP gives very inaccurate results (top), whereas our model retains the true behavior accurately (bottom).}
      \label{largeRandomErrorEst}
\end{figure}

We insert 21 large random errors at magnitude 0.1, on a randomly selected location and time ticks. Figure \ref{largeRandomErrorEst} depicts the robustness of our method.

\subsection{Local Clock Drift}
On Case300, we randomly select 5 from 22 PMUs to
create clock drift, where PMUs' internal clocks  deviate from the perfect time reference by 100 ms per second. Because local clocks are resynchronized to Coordinated Universal Time (UTC) to be coincident with a 1 pulse-per-second (pps) provided by Global Position System (GPS) \cite{Standard-PMU-IEEE}, the magnitude of the error accumulates and can become non-negligible at time samples close to the end of each re-synchronization cycle. To increase the sensitivity to detection, we defer to \eqref{eqnLossLag} and \eqref{eqnMaskLag}, learning the covariances with respect to a set of 7 lag times, roughly evenly spread out within a second: 0, 0.132, 0.264, 0.396, 0.528, 0.660, and 0.792 seconds. We set ${\beta=17}$. 

\begin{figure}[h]
    \centering
    \begin{subfigure}[h]{\linewidth}
        \centering
        \includegraphics[width=1\linewidth]{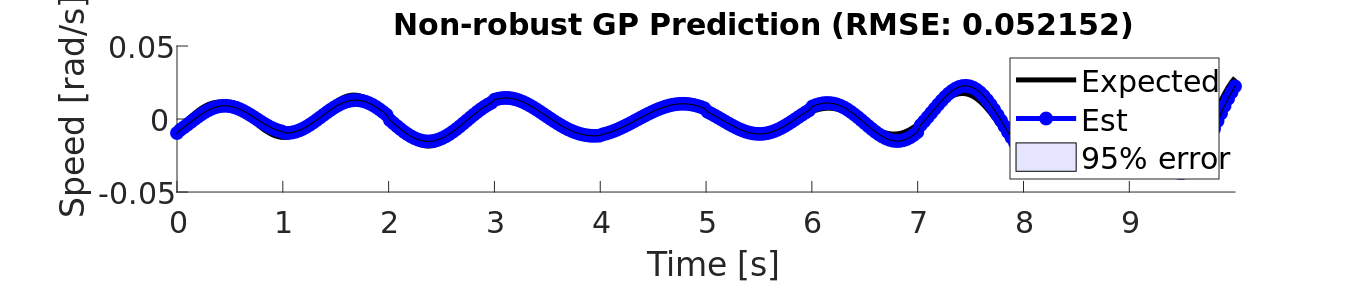}
    \end{subfigure}
    \begin{subfigure}[h]{\linewidth}
        \centering
        \includegraphics[width=1\linewidth]{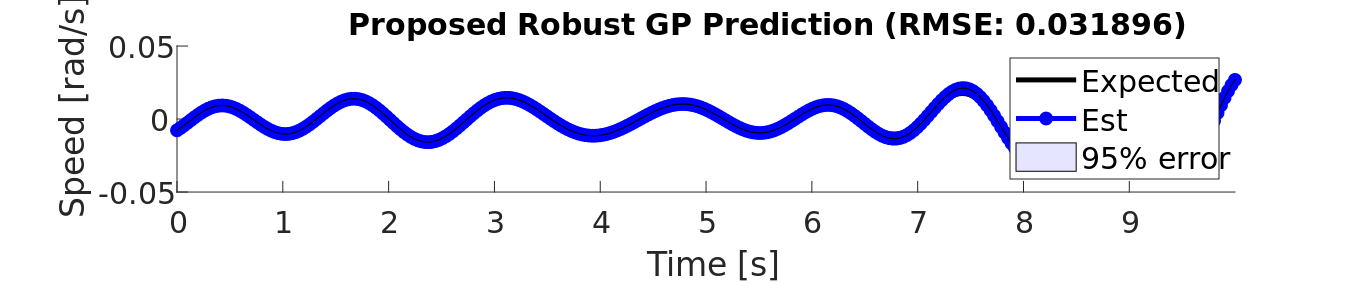}
    \end{subfigure}
    \begin{minipage}[b]{0.45\linewidth}
        \centering
        \includegraphics[width=1.1\linewidth]{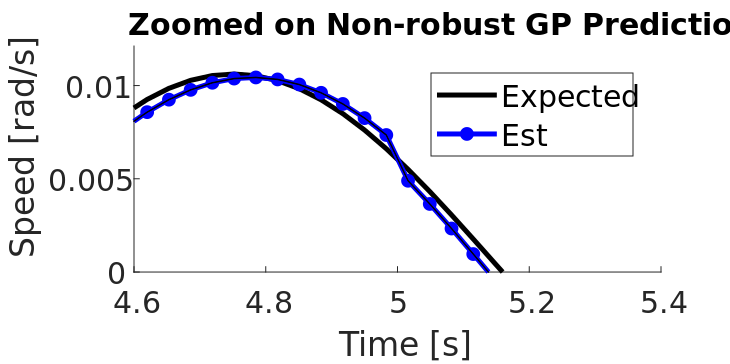}
    \end{minipage}
    \begin{minipage}[b]{0.45\linewidth}
        \centering
        \includegraphics[width=1.1\linewidth]{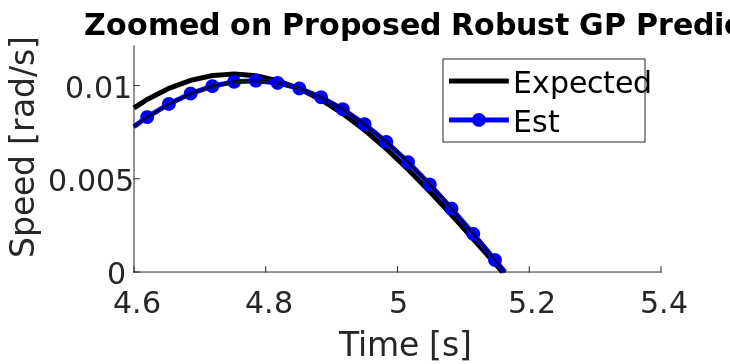}
    \end{minipage}
    \caption{{300-bus case with local clock drift}: Non-metered machine is estimated in the presence of local clock drift. Estimate from least-squares method is distorted (top, bottom left), whereas estimate from sparsity-exploiting method using L1-norm and mask is robust (middle, bottom right).}
    \label{clockDriftEst}
\end{figure}

Figure \ref{clockDriftEst} shows the results. The bottom two plots highlight an abrupt jump in the signal estimated from the traditional GP method, indicating that its inference is distorted by the drift and recovery at a re-synchronizing time point.

\subsection{Application to Large Systems}
 On case1354-pegase, clustering-based dimension reduction (DR) alleviates the computational burden of taking the inverse of a ${22339 \times 22339}$ matrix for all non-metered machines to taking the inverse of a set of matrices which, at worst, is ${5773 \times 5773}$ and, at best, is ${502 \times 502}$. The further aggregate representation (AR) of each group reduces the number of inferences from 260 non-metered machines to 26 equivalents. Table \ref{largeSysTime} highlights the speed-up.

\begin{table}[ht]
\centering
\caption{inference time of non-metered generators}
\label{largeSysTime}
\begin{tabularx}{0.8\linewidth}{ccc}
\toprule
{\bf Original} &
{\bf Dim. Red.} &
{\bf Dim. Red. + Agg. Rep.}
\\
\midrule
9895 sec & 
545 sec &
24 sec
 \\
\bottomrule
\end{tabularx}
\end{table}

\begin{figure}[h]
     \centering
     \begin{subfigure}[h]{\linewidth}
         \centering
         \includegraphics[width=1\linewidth]{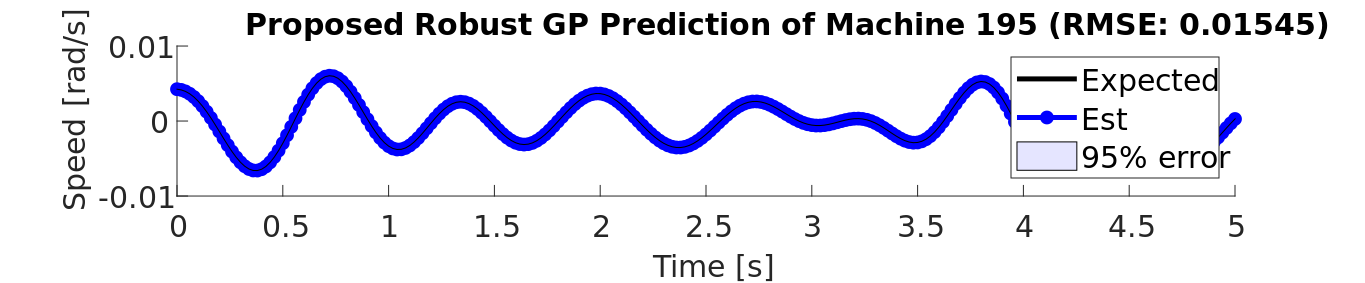}
     \end{subfigure}
     \begin{subfigure}[h]{\linewidth}
         \centering
         \includegraphics[width=1\linewidth]{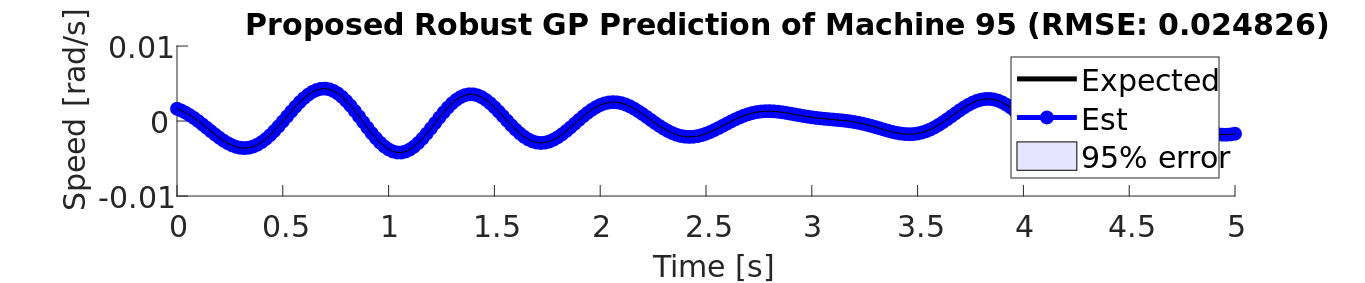}
     \end{subfigure}
     \begin{subfigure}[h]{\linewidth}
         \centering
         \includegraphics[width=1\linewidth]{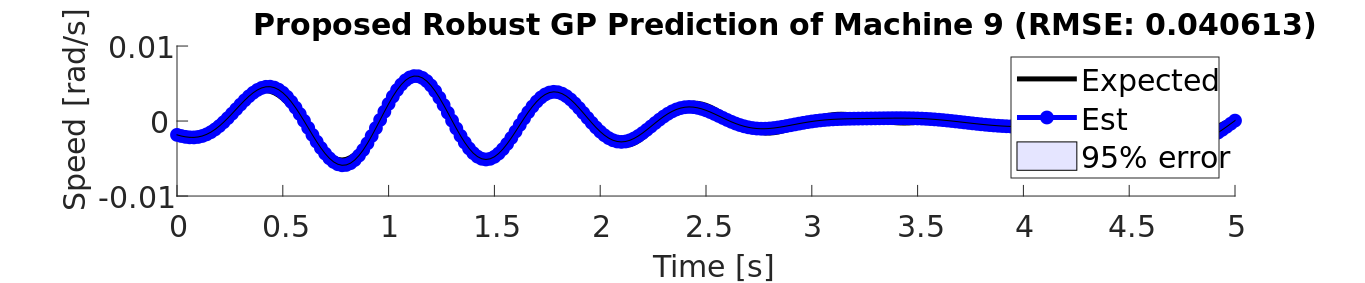}
      \end{subfigure}
      \caption{{1354-bus case with false data injection}: Non-metered locations are predicted with clustering-based dimension reduction. To assess accuracy, 3 locations corresponding to normalized RMSEs at lower-, mid-, and upper-quartile values are displayed.}
      \label{largeSysEst}
\end{figure}

\begin{figure}[h]
     \centering
     \begin{subfigure}[h]{\linewidth}
         \centering
         \includegraphics[width=1\linewidth]{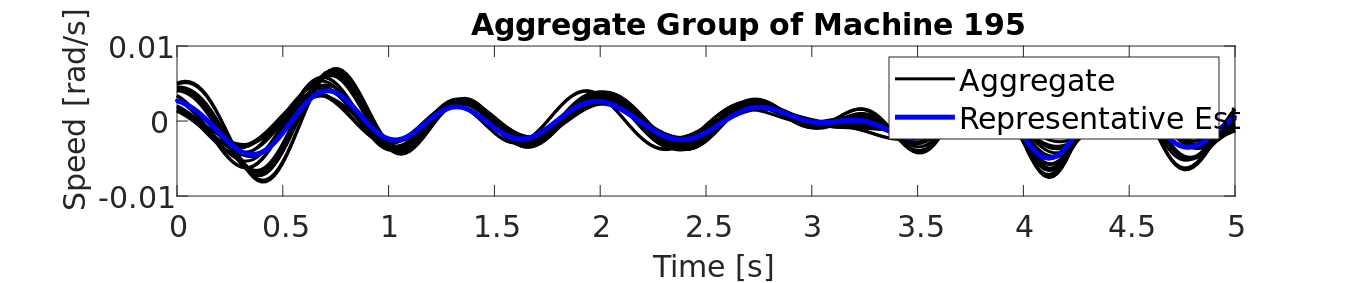}
     \end{subfigure}
     \begin{subfigure}[h]{\linewidth}
         \centering
         \includegraphics[width=1\linewidth]{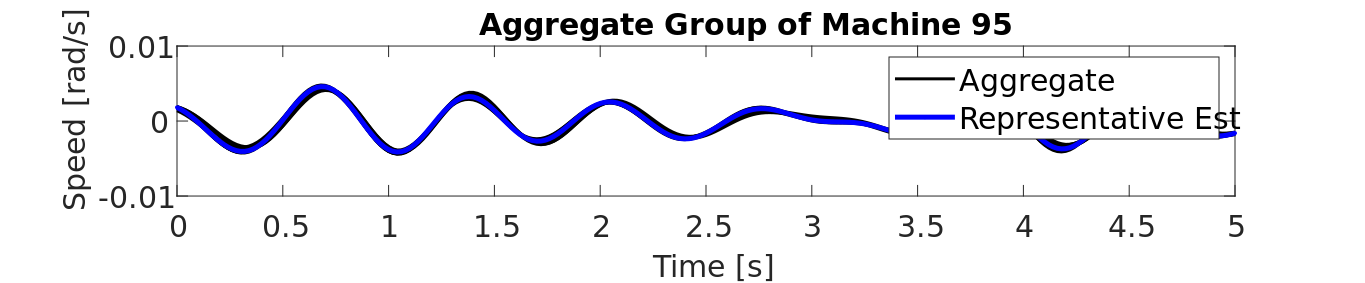}
     \end{subfigure}
     \begin{subfigure}[h]{\linewidth}
         \centering
         \includegraphics[width=1\linewidth]{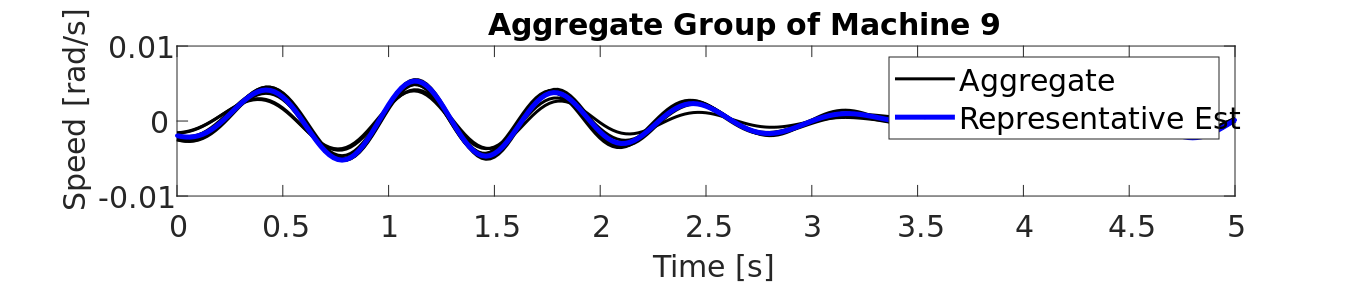}
      \end{subfigure}
      \caption{Equivalent representative estimates of groups corresponding to non-metered machines with normalized RMSEs at lower-, mid-, and upper-quartile values.}
      \label{largeSysCluster}
\end{figure}

To test the accuracy of inference using reduced ${\Sigma_{11}}$, we selectively observe the results of non-metered machines with normalized RMSEs that correspond to lower, mid, and upper quartile values among all non-metered machines. Figure \ref{largeSysEst} attests to the overall accuracy of inference with reduced ${\Sigma_{11}}$s. Figure \ref{largeSysCluster} further shows that each aggregate representative aligns well with corresponding cluster behaviors, suggesting that clustering-based heuristics are promising in GP inference.

\section{Conclusion}
\label{sec:Conclusion}

This paper presents a two-step robust Gaussian process (GP) method to robustly learn and infer grid dynamics in the presence of corrupt data. First, we adopt sparse optimization to make the learned GP model immune to a sparse distribution of data errors. Second, we learn sparse weights to identify and discard corrupted locations. We also introduce clustering-based heuristics, inspired by coherency identification, to reduce the size of the covariance matrix, facilitating fast inference on large systems.


\bibliographystyle{IEEEtran}
\bibliography{refbib}

\end{document}